\documentclass[aps,prl,showpacs,twocolumn,floatfix,amssymb,amsmath,secnumarabic,nofootinbib]{revtex4-1}
\usepackage{graphicx}
\usepackage{amsmath}
\usepackage{color}
\usepackage{threeparttable}
\begin{document}

\def\deps{\Delta\epsilon_g}
\def\tdeps{\Delta\tilde\epsilon_g}
\def\PBE{^{\rm PBE}}
\def\n{n}
\def\app{^{\rm app}}
\def\bea{\begin{eqnarray}}
\def\eea{\end{eqnarray}}
\def\ben{\begin{equation}}
\def\een{\end{equation}}
\def\sss{\scriptscriptstyle\rm}
\def\s{_{\sss S}}
\def\x{_{\sss X}}
\def\c{_{\sss C}}
\def\s{_{\sss S}}
\def\xc{_{\sss XC}}
\def\dc{_{\sss DC}}
\def\ext{_{\rm ext}}
\def\ee{_{\rm ee}}
\def\sint{ {\int d^3 r \,}}
\def\br{{\bf r}}

\title{Understanding and reducing errors in density functional calculations}
\author{Min-Cheol Kim}
\affiliation{Department of Chemistry and Institute of Nano-Bio Molecular Assemblies, 
Yonsei University, 50 Yonsei-ro Seodaemun-gu, Seoul 120-749 Korea}
\author{Eunji Sim$^*$}
\affiliation{Department of Chemistry and Institute of Nano-Bio Molecular Assemblies, 
Yonsei University, 50 Yonsei-ro Seodaemun-gu, Seoul 120-749 Korea}

\author{Kieron Burke}
\affiliation{Department of Chemistry, University of California, Irvine, CA, 92697, USA}
\date{\today}
%\sf

\begin{abstract}

We decompose the energy error of any variational
DFT calculation into a contribution due
to the approximate functional and that due to the approximate density.
Typically, the functional error dominates,
but in many interesting situations, the density-driven error dominates.
Examples range from calculations of electron affinities to
preferred geometries of
ions and radicals in solution.
In these abnormal cases, the DFT error can
be greatly reduced by using a 
more accurate density.
A small orbital gap often indicates
a substantial density-driven error.

\end{abstract}

\maketitle

Density functional theory (DFT) began with the Thomas-Fermi
approximation\cite{T27, F27}, which is now used in many branches of physics\cite{Sa91}.
For electronic structure, the
Kohn-Sham scheme\cite{KS65} is now applied 
in many other disciplines, from chemistry to materials science,
and beyond.  In all practical calculations, some form
of density functional approximation is used, leading to errors
in the property being calculated.   A persistent weakness of the
method has been an inability to control these errors
or systematically improve approximations\cite{B12}.
There are no error bars on DFT energies.
Traditionally, all that can be used
to judge the reliability of a calculation \cite{ChemGuideDFT} is 
experience with specific classes of systems and properties.

Many researchers worldwide are focused on improving
approximations to the ground-state energy functional,
but no such improvements are reported here.
On the contrary, we introduce a
general method for analyzing the error in {\em any} 
such approximate DFT calculation.
We find the somewhat surprising result that entire classes
of errors are often misclassified.
We also show how such errors can often be greatly reduced 
with relatively little computational cost.
We demonstrate the power of our method by curing
the infamous
self-interaction error (SIE) that bedevils DFT calculations of
ions and radicals in solution.
We illustrate with numerous examples from the
chemical literature, but our reasoning applies
to approximate DFT calculations in  {\em any} situation.

In ground-state DFT, the energy is written
as
\ben
E = \min_\n \left\{ F[\n] + \int d^3r\, \n(\br)\, v(\br) \right\}
\een
where $v(\br)$ is the one-body potential of the system
(e.g., the sum of attractions to the nuclei), while $F[\n]$
is a functional\cite{HK64} of the one-electron density $\n(\br)$
and
is independent of $v(\br)$.  In practical DFT calculations,
$F[\n]$ is approximated,
call it $\tilde F[\n]$.  The minimizing
density $\tilde\n(\br)$ is therefore also approximate,
so the energy error is
\ben
\Delta E = \tilde E - E = \Delta E_F + \Delta E_D
\label{DDE}
\een
where $\Delta E_F=\tilde F[\n]-F[\n]$ is
the functional error, because it is the error made by the
functional on $\n(\br)$.
The density-driven error is due to the error
in $\tilde\n(\br)$, and
$\Delta E_D$
is defined by Eq. (\ref{DDE}).

A Thomas-Fermi (TF) calculation is a pure DFT calculation, in which
$F[\n]$ itself is approximated.  Because of the inability
to treat quantum oscillations leading to shell-structure,
the density-driven
error dominates and is typically much
larger than $\Delta E_F$. 
Modern attempts at pure DFT approximations (a.k.a. orbital-free
DFT\cite{GBO07, SW09, ZL07}) are rarely tested self-consistently, for precisely this
reason.
Modern calculations employ the KS scheme
in which only a small fraction of $F[\n]$ is approximated,
the so-called exchange-correlation contribution, $E\xc[\n]$.  Even
with the simple local density approximation (LDA),
$\Delta E_F$ usually dominates over $\Delta E_D$.
Densities are often so accurate that it is common practice
to test a new approximation with orbitals from a less accurate one\cite{GJPF92,S92, OB94}.  We denote such calculations
as {\em normal}, as their energetic errors largely reflect the
true error in the approximation.

But in a small fraction of 
calculations, $\Delta E_D$ dominates over $\Delta E_F$.
In such {\em abnormal} calculations the typical
error of a given approximation appears abnormally large.
Our analysis shows that this is a qualitatively different (and more
insidious) error, due to an unusual sensitivity to the
XC potential, leading to a poor-quality density.   
Such errors should not be directly attributed to the given approximation,
but rather to the type of calculation, and
can be greatly reduced by using more accurate densities,
sometimes at little additional cost.

The infamous SIE \cite{PZ81}
made by standard DFT approximations is
``well-known'' to be extreme when an extra electron is 
added to a neutral atom or molecule.   In LDA,
H$^-$ is unbound because of this\cite{SRZ77}.
But recently\cite{LFB10} it has been shown that, if more accurate
densities are used instead of self-consistent densities, the
errors are reduced so much that they are {\em less} than
those of ionization potentials.  The SIE is reduced by adding an extra electron!

We begin with pure DFT, such as TF
calculations for total atomic energies.  
For Ra ($Z=88$),
the TF error is about -3.4 kH out of -23 kH, where kH is a kiloHartree,
and the relative error vanishes as $Z\to\infty$ \cite{LS73}.
But $\Delta E_F$ is only -0.62 kH, and so $\Delta E_D$ is about 4 times
larger.  The errors in self-consistent TF atomic calculations are
mainly due to the error in the density, and the main aim of orbital-free
approaches should be to reduce this error.

But most modern calculations use the KS scheme, solving self-consistently
\ben
\left\{- \nabla^2/2 + v\s(\br)\right\}\phi_i (\br)
=\epsilon_i\, \phi_i(\br),
\een
where $\phi_i(\br)$ is a KS orbital and $\epsilon_i$ its eigenvalue. Here,
 the density of the orbitals is defined to match the true density, and
the energy can be found from
\ben
F[\n]=T\s[\n]+U[\n]+E\xc[\n],
\een
where $T\s$ is the kinetic energy of the orbitals, $U$ their Hartree energy,
and 
\ben
v\s(\br)=v(\br)+\int d^3r \frac{\n(\br)}{|\br-\br'|} + v\xc(\br),
~v\xc(\br)= \frac{\delta E\xc}{\delta \n(\br)}.
\label{vxcdef}
\een
Approximations in common use are LDA\cite{KS65}
the generalized gradient approximation (GGA), such as PBE
\cite{PBE96}, and hybrid functionals\cite{B93, SDCF94}.  
The energy-functional error is transmitted to
the density via $v\xc(\br)$.  The SIE of
standard approximations causes
$v\xc(\br)$ to decay too rapidly with $r$, so that
$v\s(\br)$ is too shallow\cite{AV85}, and the $\epsilon_i$
are insufficiently deep by several eV. 
However, an almost constant shift in $v\s(\br)$ has 
little effect on $\tilde\n(\br)$ and therefore on $E$.

\begin{figure}[htb]
\begin{center}
\includegraphics[width=3in]{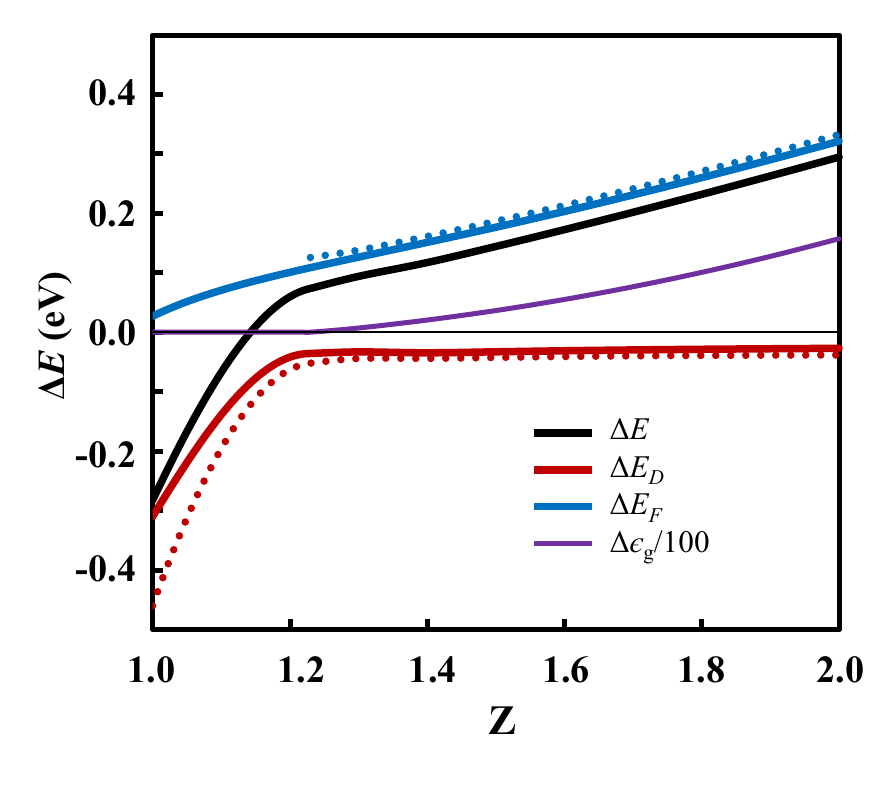}
\vskip -0.5cm
\caption{Errors in ground-state energies of two-electron ions as a function of nuclear
charge: PBE energies evaluated on exact\cite{UG94,HU97} (solid) and Hartree-Fock 
(HF, dotted) densities\cite{interpolation}.}
\label{2ePBEfig}
\end{center}
~
\vskip -1cm
\end{figure}
To illustrate our method, we apply it to 
the simplest non-trivial system,
two electron ions, with nuclear charge $Z$ varying down to
$1$ (H$^-$).  For He (and any $Z \geq 2$) with standard approximations,
$\tilde\n(r)$ is indistinguishable from $\n(r)$, despite the
large errors in $v\s(r)$ and $\epsilon_i$.  Thus $\Delta E_F$ is 0.3 eV,
while $\Delta E_D$ is only -0.04 eV, and the calculation
is {\em normal}.
But the energy error for $Z \leq 2$ in Fig. \ref{2ePBEfig}
behaves
rather smoothly until around $Z_c \approx 1.23$, where it suddenly
changes behavior.  
As $Z$ is reduced from 2 (He) to 1 (H$^-$),
a fraction of an electron\cite{SRZ77} unbinds (about 0.3)
in a standard DFT calculation\cite{PBE96}, greatly increasing the error.

Our analysis explains the origin of this error in general terms.  The solid
colored lines decompose the error into $\Delta E_F$ and $\Delta E_D$.
Around $Z_c \approx 1.23$, where $\epsilon_{1s}$ 
vanishes and the system begins to ionize, $\Delta E_D$ grows
and leads to the qualitative change in $\Delta E$.  
Nothing special happens to $\Delta E_F$ which
is almost zero for H$^-$ and is far less than for He.
A DFT calculation with an accurate two-electron density produces a smaller error for the electron
affinity of H than for the ionization energy of He\cite{LFB10}.

\begin{figure}[htb]
\begin{center}
\includegraphics[width=3in]{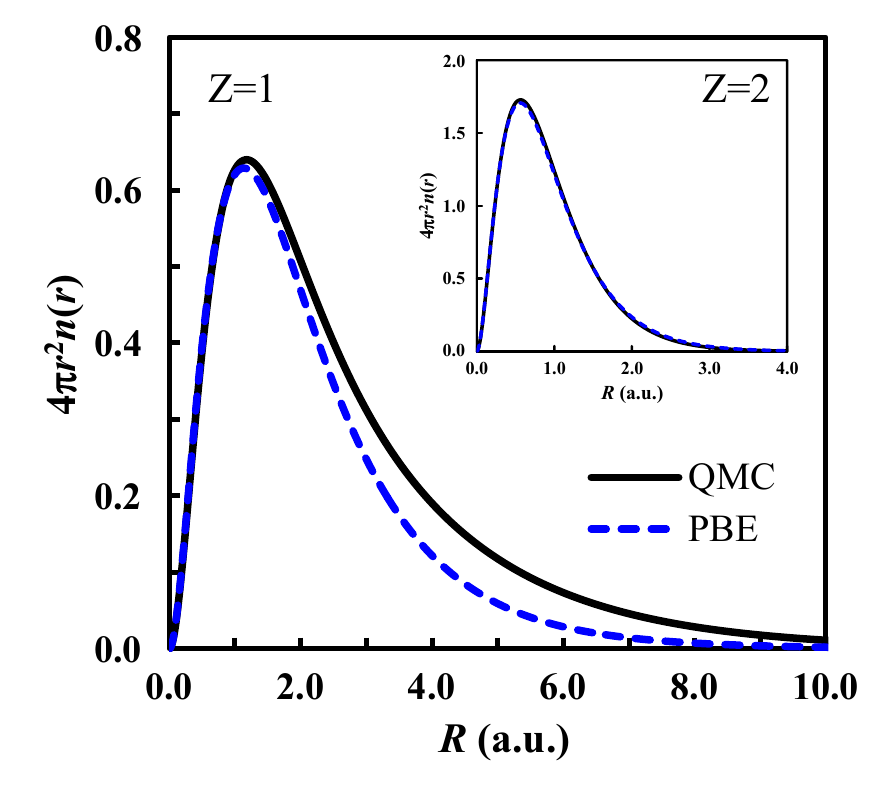}
\vskip -0.5cm
\caption{Exact and PBE radial densities for H$^{-}$ and He.}
\label{H-}
\end{center}
\end{figure}
\vskip -0.5cm

In an {\em abnormal} calculation, the system is peculiarly sensitive to the error
in $v\xc(\br)$, so that $\tilde\n(\br)$ differs significantly from $\n(\br)$,
enhancing $\Delta E_D$.
The large error in density is very visible in Fig. \ref{H-}, where the
PBE density integrates to only 1.7 electrons.
From Fig. \ref{2ePBEfig}, one sees that PBE is (somewhat
accidentally) almost exact for H$^-$ when evaluated on the exact density.

Our method can be applied to any small system where accurate densities can be
calculated via quantum chemical methods, and it will show when an error is density
driven.  But much of the value of DFT is in its relatively low computational cost,
allowing large systems to be treated, where highly accurate
densities are prohibitively expensive.  
However, if we apply linear response theory to the
KS system:
\ben
\delta \n(\br) = \int d^3r' \chi\s(\br,\br')\, \delta v\s(\br')
\een
where $\delta \n(\br)$ is the change in density induced by $\delta v\s(\br)$,
\ben
\chi\s(\br,\br')=\sum_{i,j} (f_i-f_j) 
\frac{\phi_i^*(\br)\phi_j^*(\br)\phi_i(\br')\phi_j(\br')}{\epsilon_i-\epsilon_j + i0_+}
\een
is the static density-density KS response function, 
and $f_i$ is the KS orbital
occupation factor\cite{DG90}.  The smallest denominator is $\deps$,
the HOMO-LUMO gap.
Normally, the difference between the exact and approximate $v\s(\br)$
is small, ignoring any constant shift. If
$\tdeps$ is not unusually small, this error leads
to a small error in $\tilde\n(\br)$.
But if $\tdeps$ is small, even a small error in 
$v\s(\br)$ can produce a large change in the density, and self-consistency
only increases this effect.  Thus small $\tdeps$ suggests a large density error,
and we plot $\deps\PBE$
in Fig. \ref{2ePBEfig}.
For two-electron ions, the PBE LUMO is unbound, 
so that $\deps\PBE=|\epsilon_{HOMO}\PBE|$.  At $Z_c$ this vanishes.
For other atomic anions with standard approximations, $\tilde\epsilon_{HOMO} > 0$, 
i.e., a resonance\cite{LFB10}.  Finite atom-centered basis sets
turn
this resonance into an eigenstate with an accurate density,
and produce accurate
electron affinities\cite{RTS02}.

Also, we need only a more accurate density than that the poor density of
the abnormal DFT calculation itself.
For SIE, we know that
most of the error in $v\s(\br)$ can
be cured with orbital-dependent functionals\cite{PZ81, DCM05, VVS05},
and
a the Hartree-Fock density is often sufficient and
is available in all quantum chemical codes.
Thus, the SIE density-driven error of standard approximations will
often be cured by evaluating 
DFT energies on HF densities, called HF-DFT\cite{LFB10},
which are not much more expensive than self-consistent DFT calculations.
This method yields extremely small errors (about 0.05 eV) for the
electron affinities of atoms and small molecules\cite{LFB10, KSB11}.
Occasionally, spin-contamination makes HF yield a poor density, and
so HF-DFT fails.

\begin{figure}[htb]
\begin{center}
\includegraphics[width=3in]{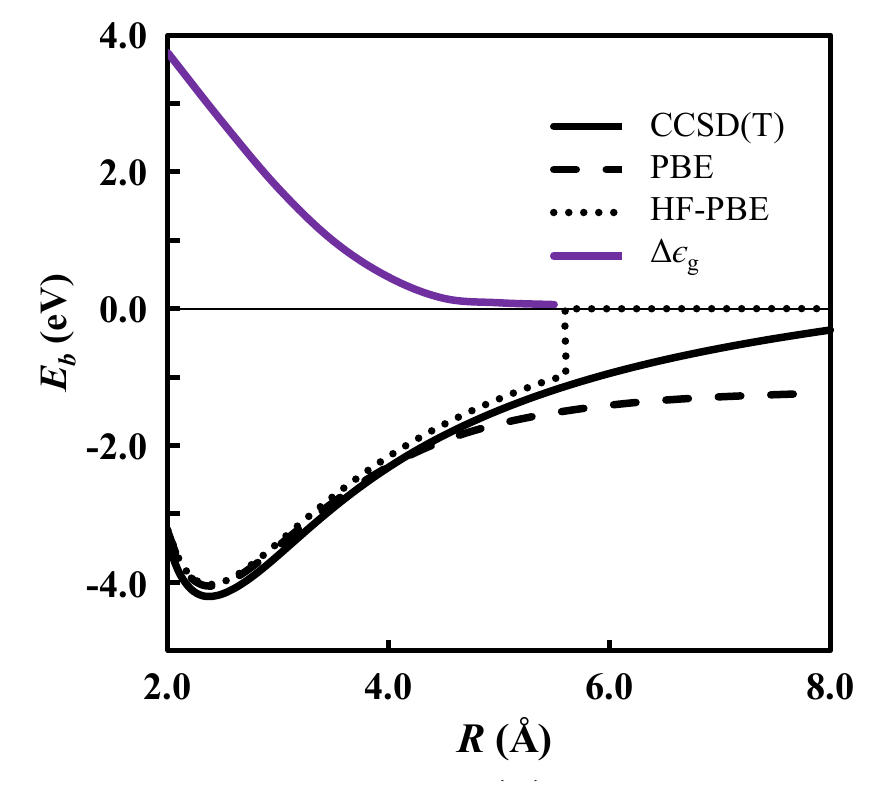}
\vskip -0.5 cm
\caption{Energy of NaCl as a function of Na-Cl distance in several calculations,
and the PBE HOMO-LUMO gap.} 
\label{NaCldissfig}
\end{center}
\vskip -0.8 cm
\end{figure}

Our next abnormality is well-known\cite{RPC06}.
DFT calculations of molecular dissociation energies ($E_{b}$) are
usefully accurate with GGA's, and more so with hybrid functionals.
These  errors are often about 0.1 eV/bond\cite{ES99}, found  by subtracting
the calculated molecular energy at its minimum from the
sum of calculated atomic energies.  This is because,
if one simply increases
the bond lengths to very large values, the fragments fail to
dissociate into neutral atoms.  
The prototypical case is NaCl, which dissociates into
Na$^{0.4}$ and Cl$^{-0.4}$ in a PBE calculation\cite{RPC06}.   
The large error in density for the stretched bond yields $\Delta E_{b}
\approx 1$ eV,
as shown as the difference between PBE and HF-PBE in Fig. \ref{NaCldissfig}.
In this case, the HF density spontaneously suddenly switches to neutral
atoms at about 5.6\AA, but is correct in the dissociation limit.
The common practice of using isolated atomic
calculations is inconsistent, but removes the density-driven error,
because isolated atoms are normal.
Incorrect dissociation occurs whenever the approximate HOMO of one is below
the LUMO of the other\cite{RPC06}, which guarantees a vanishing $\tdeps$
when the bond is greatly stretched.   The exact $v\xc(\br)$ contains
a step between the atoms which
is missed by semi-local approximations.

When modern functionals were first being adopted for molecular calculations,
they were sometimes evaluated on HF densities\cite{GJPF92,S92, OB94},
so as to compare only functional errors.
More recently, Janesko and Scuseria\cite{JS08} showed this led to
substantial improvement in transition-state barriers.
The prototype of such barriers is the symmetric H$-$H$_2$ transition state, 
which is improved by almost a factor of 2 by using HF-PBE instead of PBE.  
Here $\deps\PBE$ is not quite as small (2.5 eV) as in other cases, but
the improvement upon using the HF densities is still substantial. 
High-level {\em ab initio} calculations yield an energy barrier
of 0.43 eV\cite{JS08}, where PBE gives value of 0.16 eV and HF-PBE gives 0.25 eV.
Analysis of a collection of barriers in Table 1 of Ref. \onlinecite{VPB12} shows that,
in cases where the HF-DFT barrier differs from the self-consistent barrier by more than,
for instance 25\%, the mean absolute error is more than three
times smaller than DFT.   The sole exception is the
 t-N$_2$H$_2$ hydrogen transfer
forward reaction barrier, where the HF density
is spin-contaminated (just as in the molecule CN\cite{KSB11}).

Finally we report new
applications where we drive out
the density-driven error.
The potential energy
surfaces (PES) of odd-electron radical complexes like 
OH$\cdot$Cl$^{-}$ and OH$\cdot$H$_2$O are
important in radiation,
atmospheric, and environmental chemistry, as well as 
in cell biology\cite{IntroRadiatChem,
AtmosChemPhys, POM06, RadBioMed}.
For example, how anions behave in droplets
is critical to our understanding
aerosols in the atmosphere\cite{JPS73}.  Accepted
wisdom is that anions near an 
air-water interface, being less screened, have lower concentrations\cite{JT00}. 
But recent classical molecular dynamics 
(MD) simulations have shown the opposite\cite{KLFJ00, JT00}.

This controversy invites an  {\em ab initio} MD approach,
to either reinforce or debunk classical MD.  
However, DFT approximations have problems here
\cite{GKC04, GKC04b, PZ02, VVS05}.
Several studies show 
two minima in the ground-state PES: 
a normal hydrogen bond and a
2-center, 3-electron interacting hemi-bond\cite{C11}. 
High-level quantum chemical calculations\cite{C11} and self-interaction
corrected DFT \cite{VVS05, DKT08} reveal that the true PES has only one minimum, 
the hydrogen-bonding structure.
Hemi bonding is overstabilized in
approximate DFT because three electrons
incorrectly delocalize over
two atoms.

\begin{figure}[htb]
\begin{center}
\includegraphics[width=3in]{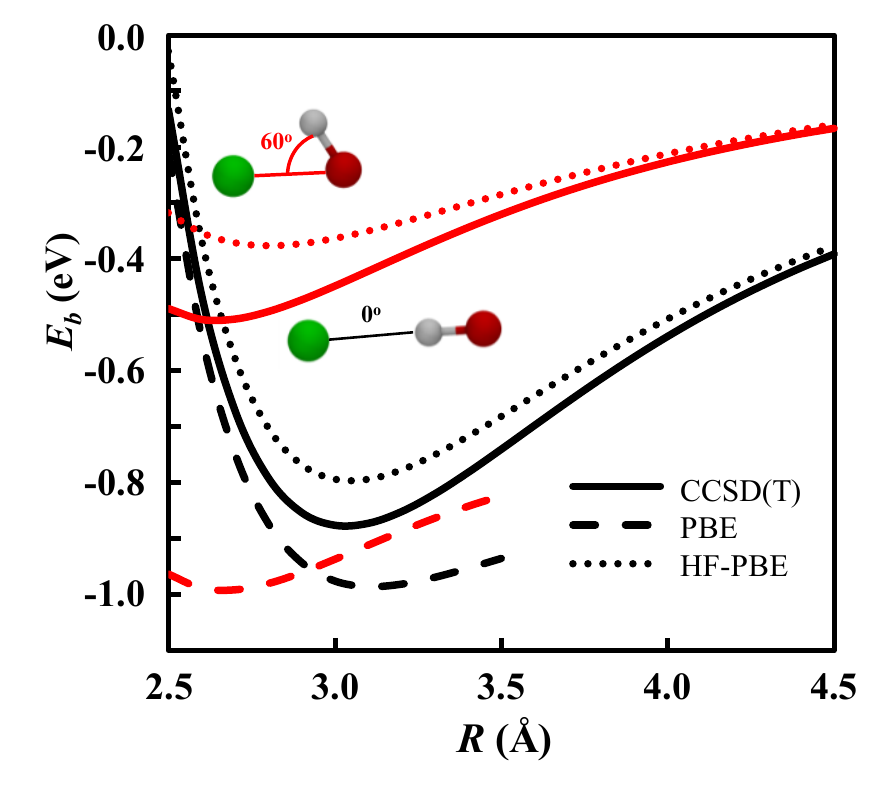}
\vskip -0.5 cm
\caption{Potential energy surface scans for the 
HO$\cdot$Cl$^{-}$ complex along Cl - O - H angle
of 0$^{\circ}$ (hydrogen-bonding structure, black lines)
and 60$^{\circ}$ (hemi-bonding structure, red lines) using various methods; $R$ is the Cl-O
separation.} 
\label{HOClPESfig}
\end{center}
~
\vskip -1 cm
\end{figure}

In Fig. \ref{HOClPESfig}, we show plots of the PES of
HO$\cdot$Cl$^-$ complex using different methods. 
The O-H bond length was fixed at 1 \AA. 
The binding energy is
\ben
\Delta E_{b}
= 
E_{\rm{HO}\cdot\rm{Cl}^{-}}(R, \theta) - ( E_{\rm{HO}\cdot} + E_{\rm{Cl}^{-}} ),
\een 
where $E_{\rm{HO} \cdot \rm{Cl} ^{-}}(R, \theta)$ is the energy on
a given geometry with Cl - O distance $R$, and Cl - O - H angle $\theta$, $E_{\rm{HO}\cdot}$
is the energy of the OH radical, and $E_{\rm{Cl}^{-}}$ is the energy of Cl anion.
The difference between the energy minima
of hydrogen- and hemi-bonding structures in PBE is less than 0.01 eV.
A small $\deps\PBE$ in the hydrogen bonding
structure ($\sim$ 0.32 eV) suggests a large density-driven error.
We find HF-PBE follows the same trends and produces the same minima as CCSD(T),
although the binding energies themselves have errors of up to 0.09 eV. 
Similar conclusions are found for the
the PES of the HO$\cdot$H$_{2}$O complex\cite{hoh2opes}. 

In every example of density-driven error in this paper, the HOMO-LUMO gap of the
DFT calculation is small.
We end with an example in which a small gap does not produce a density-driven error. Much recent research is focused on localization errors of approximations\cite{CMY08}, and many failures can be related to such errors. Consider the classic example of a severe SIE, namely stretched H$_{2}^{+}$ with a standard functional. As the bond is stretched, the gap rapidly
shrinks, suggesting abnormality, but when a HF-DFT calculation is performed, the error barely changes. Thus, this is a normal calculation whose error is functional-driven, not density-driven, and HF-DFT does not reduce the error. The small gap is due to stretching the bond, not a sign of an incipient density-driven error\cite{h2dissociation}.

Our method for classifying DFT errors is general.  Declaring a calculation
abnormal depends on both the energy being calculated (total, ionization, bond, etc.)
and the approximation
being used.  The error in {\em any} approximation can be studied in this way.
For example, methods that begin from exact exchange (such as RPA\cite{LP75, NFG03, FV05} or
ab-initio DFT\cite{BLS05}) which yield better potentials, could be examined to see
if such improvements yield better energies.
We focused here on the SIE because of its ubiquity,
but one can apply the same reasoning to, e.g., the
correlation energy itself 
\cite{KK08, GBMM12}, or the error in the KS kinetic energy in orbital-free
approximations\cite{SW09}. 
The classic examples of stretched H$_2$ and H$_2^+$
are normal, because self-consistent densities (restricted in the case of H$_2$)††
close to exact densities.  
The myriad materials and molecules where standard DFT fails should 
now be re-examined, to distinguish
between true errors (i.e., large energy errors even on 
exact densities) versus density-driven errors, which are 
system- and property-dependent.
 
Calculation details are given in Supplemental Material\cite{calculation}.
We thank Prof. Doug Tobias for insightful discussions. 
This work was supported by the global research network grant (NRF-2010-220-C00017)
and the national research foundation (2012R1A1A2004782, ES). KB acknowledges
 support under NSF CHE-1112442.

\vspace{0.5cm}\rule{\linewidth}{0.5mm}\vspace{0.2cm}
* Corresponding Author: esim@yonsei.ac.kr \vspace{0.5mm}


%\bibliography{lib3}

\end{document}